\documentclass[runningheads]{llncs}
\usepackage{graphicx}
\usepackage{comment}
\usepackage{amsmath,amssymb}
\usepackage{color}

\usepackage[disable] {easy-todo}
\usepackage{CJKutf8}
\usepackage[mathscr]{eucal}
\usepackage{bm}
\begin{document}
\begin{CJK}{UTF8}{ipxm}
\title{Order-Guided Disentangled Representation Learning for Ulcerative Colitis Classification with Limited Labels}
\author{Shota Harada\inst{1} \and
Ryoma Bise\inst{1,2} \and
Hideaki Hayashi\inst{1} \and
Kiyohito Tanaka\inst{3} \and
Seiichi Uchida\inst{1,2}}

\authorrunning{S. Harada et al.}

\institute{Kyushu University, Fukuoka City, Japan.
\email{shota.harada@human.ait.kyushu-u.ac.jp} \and
National Institute of Informatics, Tokyo, Japan. \and
Kyoto Second Red Cross Hospital, Kyoto, Japan.}
\maketitle
\begin{abstract}
Ulcerative colitis (UC) classification, which is an important task for endoscopic diagnosis, involves two main difficulties. First, endoscopic images with the annotation about UC (positive or negative) are usually limited. Second, they show a large variability in their appearance due to the location in the colon. Especially, the second difficulty prevents us from using existing semi-supervised learning techniques, which are the common remedy for the first difficulty. In this paper, we propose a practical semi-supervised learning method for UC classification by newly exploiting two additional features, the location in a colon (e.g., left colon) and image capturing order, both of which are often attached to individual images in endoscopic image sequences. The proposed method can extract the essential information of UC classification efficiently by a disentanglement process with those features. Experimental results demonstrate that the proposed method outperforms several existing semi-supervised learning methods in the classification task, even with a small number of annotated images.

\end{abstract}

\section{Introduction}

\begin{figure}[t]
\includegraphics[width=\textwidth]{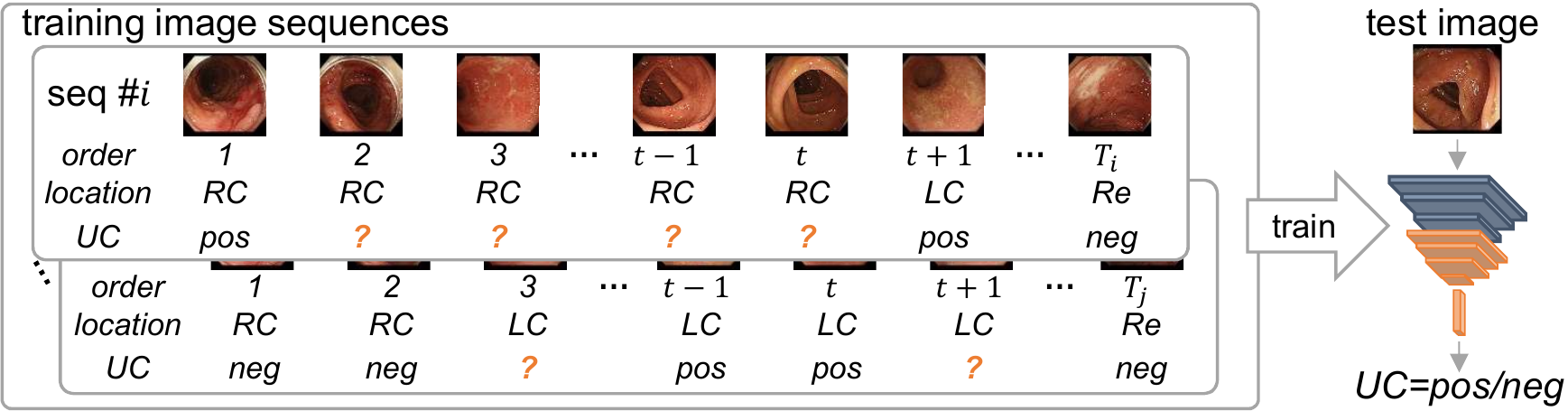}
\caption{Underlying concept for the proposed method. The objective of the study is to train an ulcerative colitis (UC) classifier with incomplete UC labels. The order and location are used as the guiding information (RC: right colon. LC: left colon. Re: rectum).}
\label{fig: overview}
\end{figure}

In the classification of ulcerative colitis (UC) using deep neural networks, where endoscopic images are classified into lesion and normal classes, it is difficult to collect a sufficient number of labeled images because the annotation requires significant effort by medical experts. UC is an inflammatory bowel disease that causes inflammation and ulcers in the colon. Specialist knowledge is required to annotate UC because texture features, such as bleeding, visible vascular patterns, and ulcers, should be captured among the image appearances that drastically vary depending on the location in the colon to detect UC.

Semi-supervised learning methods~\cite{Arazo_IJCNN2020,Wallach_NIPS2019,Lee_ICML2013,Sohn_NIPS2020} have been used to train classifiers based on a limited number of labeled images, involving the use of both labeled and unlabeled images. If a classifier with a moderate classification performance is obtained with few labeled data, the performance of a classifier can be further improved by applying these semi-supervised learning methods. However, existing semi-supervised learning methods do not show satisfactory performance for UC classification because they implicitly assume that the major appearance of images is determined by the classification target class, whereas the major appearance of UC images is determined by the location in the colon, not by the disease condition.

Incorporating domain-dependent knowledge can also compensate for the lack of labeled data. In endoscopic images, we can utilize two types of prior knowledge: location information and temporal ordering information, that is, the order in which the endoscopic images were captured. Location information can be obtained easily by tracking the movement of the endoscope during the examination~\cite{Herp2021,Mori2001}, with the rough appearance of endoscopic images characterized by their location. Endoscopic images are acquired in sequence while the endoscope is moved through the colon. Therefore, the temporal ordering information is readily available, and temporally adjacent images tend to belong to the same UC label. If the above information can be incorporated into semi-supervised learning, more accurate and reliable networks for UC classification can be developed.

In this study, we propose a semi-supervised learning method for UC classification that utilizes location and temporal ordering information obtained from endoscopic images. Fig.~\ref{fig: overview} shows the underlying concept for the proposed method. In the proposed method, a UC classifier is trained with incomplete UC labels, whereas the location and ordering information are available. By utilizing the location information, we aim to improve UC classification performance by simultaneously extracting the UC and location features from endoscopic images. We introduce disentangled representation learning~\cite{Liu_NIPS2018,Liu_CVPR2018} to effectively embed the UC and location features into the feature space separately. To compensate for the lack of UC-labeled data using temporal ordering information, we formulated the ordinal loss, which is an objective function that brings temporally adjacent images closer in the feature space. 

The contributions of this study are as follows:
\begin{itemize}
\item We propose a semi-supervised learning method that utilizes the location and temporal ordering information for UC classification. The proposed method introduces disentangled representation learning using location information to extract UC classification features that are separated from the location features.
\item We formulate an objective function for order-guided learning to utilize temporal ordering information of endoscopic images. Order-guided learning can obtain the effective feature for classifying UC from unlabeled images by considering the relationship between the temporally adjacent images. 
\end{itemize}

\section{Related work}
Semi-supervised learning methods that utilize unlabeled samples efficiently have been reported in the training of classifiers when limited labeled data are available~\cite{Arazo_IJCNN2020,Wallach_NIPS2019,Lee_ICML2013,Sohn_NIPS2020}.
Lee~\cite{Lee_ICML2013} proposed a method called Pseudo-Label, which uses the class predicted by the trained classifier as the ground-truth for unlabeled samples. Despite its simplicity, this method improves the classification performance in situations where labeled images are limited. Sohn~\textit{et al.}~\cite{Sohn_NIPS2020} proposed FixMatch, which improves the classification performance by making the predictions for weakly and strongly augmented unlabeled images closer during training. These semi-supervised learning methods work well when a classifier with a moderate classification performance has already been obtained using limited labels. However, in UC classification, which requires the learning of texture features from endoscopic images whose appearance varies depending on imaging location, it is difficult to obtain a classifier with a moderate classification performance using limited labeled endoscopic images, and applying these methods to UC classifications may not improve classification performance. Therefore, we propose a semi-supervised learning method that does not directly use the prediction results returned by a classifier trained by limited-labeled data, but utilizes two additional features: the location and the temporal ordering.

Several methods that utilize the temporal ordering information of images have been reported~\cite{Cao_2020_CVPR,Dwibedi_CVPR2019,Harada_EMBC2019}.
For example, Cao~\textit{et el.}~\cite{Cao_2020_CVPR} proposed Temporal-Cycle Consistency (TCC), which is a self-supervised learning method that utilizes temporal alignment between sequences.
The TCC yields good image feature representation by maximizing the number of points where the temporal alignment matches. Dwibedi~\textit{et al.}~\cite{Dwibedi_CVPR2019} proposed a few-shot video classification method that utilizes temporal alignment between labeled and unlabeled video, then improved the video classification accuracy by minimizing the distance between temporally aligned frames. Moreover, a method for segmenting endoscopic image sequences has been proposed~\cite{Harada_EMBC2019}. By utilizing the prior knowledge that temporally adjacent images tend to belong to the same class, this method segments an image sequence without requiring additional annotation. However, the methods proposed \cite{Cao_2020_CVPR,Dwibedi_CVPR2019} are not suitable for our task, where involves a sequence with indefinite class transitions, because they assume that the class transitions in the sequence are the same. Furthermore, the method proposed in \cite{Harada_EMBC2019}, which assumes segmentation of normal organ image sequences, is not suitable for our task where the target image sequence consists of images of both normal and inflamed organs. In the proposed method, temporal ordering information is used to implement order-guided learning, which brings together temporal adjacency images that tend to belong to the same UC class, thus obtaining a good feature representation for detecting UC in the feature space.

\section{Order-guided disentangled representation learning for UC classification with limited labels}

\begin{figure}[t]
\includegraphics[width=\textwidth]{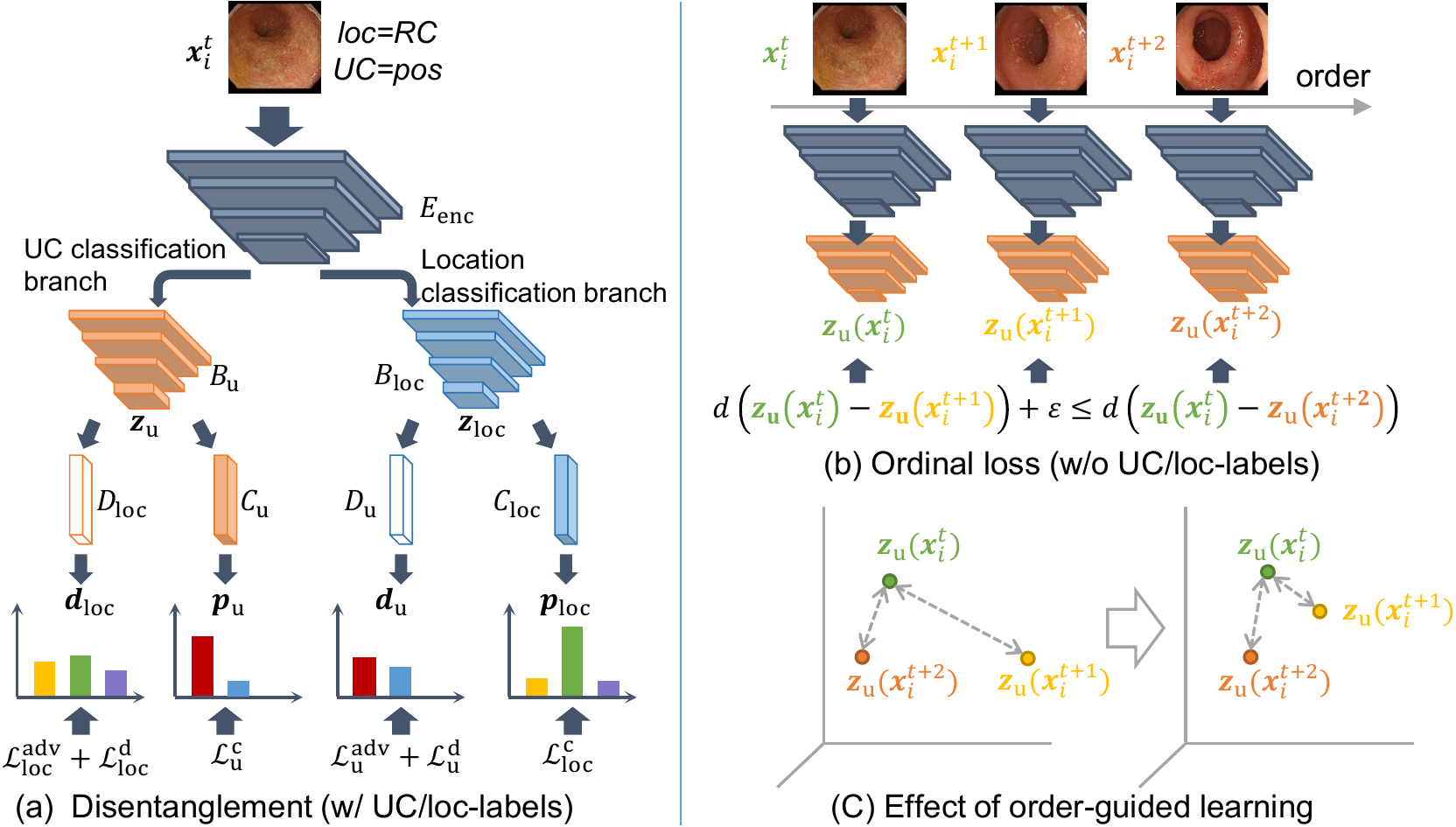}
\caption{Overview of the proposed method. (a) Disentanglement into the UC feature $\boldsymbol{z}_{\textrm{u}}$ and the location feature $\boldsymbol{z}_{\textrm{loc}}$. (b) Ordinal loss for order-guided learning. (c) Effect of order-guided learning.} 
\label{fig: prop}
\end{figure}

The classification of UC using deep neural networks trained by general learning methods is difficult for two reasons. First, the appearances of the endoscopic images vary dynamically depending on the location in the colon, whereas UC is characterized by the texture of the colon surface. Second, the number of UC-labeled images is limited because annotating UC labels to a large number of images requires significant effort by medical experts. 

To overcome these difficulties, the proposed method introduces \textit{disentangled representation learning} and \textit{order-guided learning}. Fig.~\ref{fig: prop} shows the overview of the proposed method. In disentangled representation learning using location information, we disentangle the image features into features for UC-dependent and location-dependent to mitigate the worse effect from the various appearance depending on the location. Order-guided learning utilizes the characteristics of an endoscopic image sequence in which temporally adjacent images tend to belong to the same class. We formulated an objective function that represents this characteristic and employs it during learning to address the limitation of the UC-labeled images.

\subsection{Disentangled representation learning using location information}
Disentangled representation learning for the proposed method aims to separate the image features into UC and location-dependent features. These features are obtained via multi-task learning of UC and location classification. Along with the training of classifiers for UC and location classification tasks, the feature for one task is learned to fool the classifier for the other task; that is, the UC-dependent feature is learned to be non-discriminative with respect to location classification, and vice versa.

The network structure for learning disentangled representations is shown in Fig.~\ref{fig: prop}(a). This network has a hierarchical structure in which a feature extraction module branches into two task-specific modules, each of which further branches into two classification modules. The feature extraction module $E_{\mathrm{enc}}$ extracts a common feature vector for UC and location classification from the input image. The task-specific modules $B_{\mathrm{u}}$ and $B_{\mathrm{loc}}$ extract the UC feature $\boldsymbol{z}_{\textrm{u}}$ and the location feature $\boldsymbol{z}_{\textrm{loc}}$, which are disentangled features for UC and location classification. Out of four classification modules, the modules $C_{\mathrm{u}}$ and $C_{\mathrm{loc}}$ are used for UC and location classification, respectively, whereas $D_{\mathrm{u}}$ and $D_{\mathrm{loc}}$ are used to learn the disentangled representations.

In the left branch of Fig.~\ref{fig: prop}(a), the network obtains the prediction results for UC classes, $\boldsymbol{p}_{\mathrm{u}}$, as the posterior probabilities, based on the disentangled UC feature $\boldsymbol{z}_{\mathrm{u}}$ through learning. Hereinafter, we explain only the training of the left branch in detail because that of the right branch can be formulated by simply swapping the subscripts ``loc'' and ``u'' in the symbols for the left branch.

Given a set of $N$ image sequences and corresponding location class labels $\{\bm{x}_{i}^{(1:T_i)}, \bm{l}_{i}^{(1:T_i)}\}^{N}_{i=1}$ and a set of limited UC class labels $\{\bm{u}_{j}^{k}\mid(j, k)\in \mathcal{U}\}$, where $T_i$ is the number of images in the $i$-th image sequence and $\bm{u}_{j}^{k}$ is the UC class label corresponding to the $j$-th image in the $k$-th sequence, the training is performed based on three losses: classification loss $\mathcal{L}^{\textrm{c}}_{\textrm{u}}$, discriminative loss $\mathcal{L}^{\textrm{d}}_{\textrm{loc}}$, and adversarial loss $\mathcal{L}^{\textrm{adv}}_{\textrm{loc}}$. To learn the UC classification, we minimize the classification loss $\mathcal{L}^{\textrm{c}}_{\textrm{u}}$, which is computed by taking the cross-entropy between the UC class label $\boldsymbol{u}_{i}^{t}$ and the UC class prediction $\boldsymbol{p}_{\mathrm{u}}(\boldsymbol{x}_{i}^{t})$ that is output from $C_\mathrm{u}$. The discriminative loss $\mathcal{L}^{\textrm{d}}_{\textrm{loc}}$ and adversarial loss $\mathcal{L}^{\textrm{adv}}_{\textrm{loc}}$ are used to learn the disentangled representation, and are formulated as follows:
\begin{gather}
\label{eq:adv_loss_entropy_u}
\mathcal{L}^{\textrm{d}}_{\textrm{loc}}(\boldsymbol{x}_{i}^{t}) = - \sum_{j=1}^{K_{\textrm{loc}}}l_{i}^{t}~\textrm{log}~d_{\textrm{loc}}^{j}(\boldsymbol{x}_{i}^{t}), ~
\mathcal{L}^{\textrm{adv}}_{\textrm{loc}}(\boldsymbol{x}_{i}^{t}) = \sum_{j=1}^{K_{\textrm{loc}}} \textrm{log}~d_{\textrm{loc}}^{j}(\boldsymbol{x}_{i}^{t}),
\end{gather}
where $\boldsymbol{d}_{\mathrm{loc}}(\boldsymbol{x}_{i}^{t})$ is the location class prediction estimated by $D_\mathrm{loc}$. By minimizing the discriminative loss $\mathcal{L}^{\textrm{d}}_{\textrm{loc}}$, the classification module $D_{\mathrm{loc}}$ is trained to classify the location. In contrast, the minimization of the adversarial loss $\mathcal{L}^{\textrm{adv}}_{\textrm{loc}}$ results in the UC feature $\boldsymbol{z}_{\textrm{u}}$ that is non-discriminative with respect to the location. Note that $\mathcal{L}^{\textrm{d}}_{\textrm{loc}}$ is back-propagated only to $D_{\mathrm{loc}}$, whereas the parameters of $D_{\mathrm{loc}}$ are frozen during the back-propagation of $\mathcal{L}^{\textrm{adv}}_{\textrm{loc}}$. As mentioned above, some images are not labeled for UC classification in this problem. Therefore, the classification loss $\mathcal{L}^\textrm{c}_\textrm{u}$ and the disentangle losses $\mathcal{L}^\textrm{adv}_\textrm{u}$ and $ \mathcal{L}^\textrm{d}_\textrm{u}$ are ignored for UC-unlabeled images.

\subsection{Order-guided learning}
Order-guided learning considers the relationship between temporally adjacent images, as shown in Fig.~\ref{fig: prop}(b). Since an endoscopic image is more likely to belong to the same UC class as its temporally adjacent images than the UC class of temporally distant images, the UC-dependent features of temporally adjacent images should be close to each other. To incorporate this assumption into learning of the network, the ordinal loss for order-guided learning is formulated as:

\begin{flalign}
\label{eq:triplet}
\mathcal{L}_{\textrm{seq}}(\boldsymbol{x}_{i}^{t}, \boldsymbol{x}_{i}^{t+1}, \boldsymbol{x}_{i}^{t+2})\!=\!\left[ ||\boldsymbol{z}_{\textrm{u}}(\boldsymbol{x}_{i}^{t})\!-\!\boldsymbol{z}_{\textrm{u}}(\boldsymbol{x}_{i}^{t+1})||^2_2\!-\!||\boldsymbol{z}_{\textrm{u}}(\boldsymbol{x}_{i}^{t})-\boldsymbol{z}_{\textrm{u}}(\boldsymbol{x}_{i}^{t+2})||^2_2\!+\! \varepsilon \right]_{+}, 
\end{flalign}
where $\boldsymbol{z}_{\textrm{u}}(\boldsymbol{x}_{i}^{t})$ is a UC feature vector for the sample $\boldsymbol{x}_{i}^{t}$ and is extracted via $E_{\mathrm{enc}}$ and $B_{\mathrm{u}}$, $[\cdot]_+$ is a function that returns zero for a negative input and outputs the input directly otherwise, and $\varepsilon$ is a margin that controls the degree of discrepancy between two temporally separated samples.\par

The UC features of temporally adjacent samples get closer by updating the network with the order-guided learning, as shown in Fig.~\ref{fig: prop}(c). This warping in the UC feature space functions as a regularization that allows the network to make more correct predictions because the temporally adjacent images tend to belong to the same UC class. The order-guided learning can be applied without the UC label, and therefore it is also effective for the UC-unlabeled images.

\section{Experimental results}
We conducted the UC classification experiment to evaluate the validity of the proposed method. In the experiment, we used an endoscopic image dataset collected from the Kyoto Second Red Cross Hospital. Participating patients were informed of the aim of the study and provided written informed consent before participating in the trial. The experiment was approved by the Ethics Committee of the Kyoto Second Red Cross Hospital.

\subsection{Dataset}
The dataset consists of $388$ endoscopic image sequences, each of which contains a different number of images, comprising $10{,}262$ images in total. UC and location labels were attached to each image based on annotations by medical experts. Out of $10{,}262$ images, $6{,}678$ were labeled as UC (positive) and the remaining $3{,}584$ were normal (negative). There were three classes for the location label: right colon, left colon, and rectum. In the experiments, the dataset was randomly split into image sequence units, and $7{,}183$, $2{,}052$, and $1{,}027$ images were used as training, validation, and test set, respectively. To simulate the limitation of the UC-labeled images, the labeled image ratio $R$ for the training set used by the semi-supervised learning methods was set to $0.1$.\par

\subsection{Experimental conditions}
We compared the proposed method with two semi-supervised learning methods. One is the Pseudo-Label~\cite{Lee_ICML2013}, which is one of the famous semi-supervised learning methods. The other is FixMatch~\cite{Sohn_NIPS2020}, which is the state-of-the-art semi-supervised learning method for the general image classification task. Since the distribution of data differs greatly between general and endoscopic images, we changed the details of FixMach to maximize its performance for UC classification. Specifically, strong augmentation was changed to weak augmentation, and weak augmentation was changed to rotation-only augmentation for processing unlabeled images. We also compared the proposed method with two classifiers trained with only labeled images in the training set with the labeled image ratio $R = 0.1$ and $1.0$.\par

In addition, we conducted an ablation study to evaluate the effectiveness of the location label, disentangled representation learning, and order-guided learning. The best network parameter for each method was determined based on the accuracy of the validation set. We used precision, recall, F1 score, specificity, and accuracy as the performance measures.

\subsection{Results}
Table~\ref{tab: compare} shows the result of the quantitative performance evaluation for each method. Excluding specificity, the proposed method achieved the best performance for all performance measures. Although the specificity of the proposed method was the third-best, it was hardly different from that of the fully supervised classification. Moreover, we confirmed that the proposed method improved all measures of the classifier trained using  only UC-labeled images in the training set with $R = 0.1$. In particular, the improvement in recall was confirmed only in the proposed method. Therefore, disentangled representation learning and order-guided learning, which use additional information other than UC information, were effective for improving UC classification performance.\par 
\begin{table}[t]
\centering
\caption{Quantitative performance evaluation. Labeled image ratio $R$ represents the ratio of the UC-labeled images in the training set.}
\label{tab: compare}
\begin{tabular}[t]{l c c c c c c}
Method & $R$ & Precision & Recall & F1 & Specificity & Accuracy\\ \hline
Supervised learning & $1.0$ & $80.52$ & $84.89$ & $82.64$ & $90.23$ & $88.51$ \\
\cline{2-7}
 & $0.1$ & $69.16$ & $67.07$ & $68.10$ & $85.78$ & $79.75$ \\
Pseudo-Label~\cite{Lee_ICML2013} & $0.1$ & $75.19$ & $61.33$ & $67.55$ & $90.37$ & $81.10$\\
FixMatch~\cite{Sohn_NIPS2020} & $0.1$ & $75.24$ & $46.82$ & $57.73$ & $\mathbf{92.67}$ & $77.90$\\
Proposed & $0.1$ & $\mathbf{77.56}$ & $\mathbf{73.11}$ & $\mathbf{75.27}$ & $89.94$ & $\mathbf{84.52}$ \\
\hline
\end{tabular}
\end{table}

\begin{table}[t]
\caption{Results of the ablation study with the location label (Location), and disentangled representation learning (Disentangle), and order-guided learning(Order)}
\label{tab: ablation}
\centering
\begin{tabular}[t]{c c c | c c c c c}
Location & Disentangle & Order & Precision & Recall & F1 & Specificity & Accuracy \\ \hline
& & & $69.16$ & $67.07$ & $68.10$ & $85.78$ & $79.75$ \\
\checkmark & & & $\mathbf{79.50}$ & $67.98$ & $73.29$ & $\mathbf{91.67}$ & $84.03$ \\
\checkmark & \checkmark & & $72.02$ & $\mathbf{73.11}$ & $72.56$ & $86.49$ & $82.18$ \\
\checkmark & \checkmark & \checkmark & $77.56$ & $\mathbf{73.11}$ & $\mathbf{75.27}$ & $89.94$ & $\mathbf{84.52}$ \\
\hline
\end{tabular}
\end{table}
Table~\ref{tab: ablation} shows the results of the ablation study. The results demonstrated that each element of the proposed method was effective for improving the UC classification. The location information was effective for improving the precision with keeping the recall. Moreover, since the recall and the specificity were improved using the order-guided learning, temporal ordering information was useful for incorporating the order-related feature that cannot be learned by only the annotation to individual images.\par 

\begin{figure}[t]
\centering
\includegraphics[width=\textwidth]{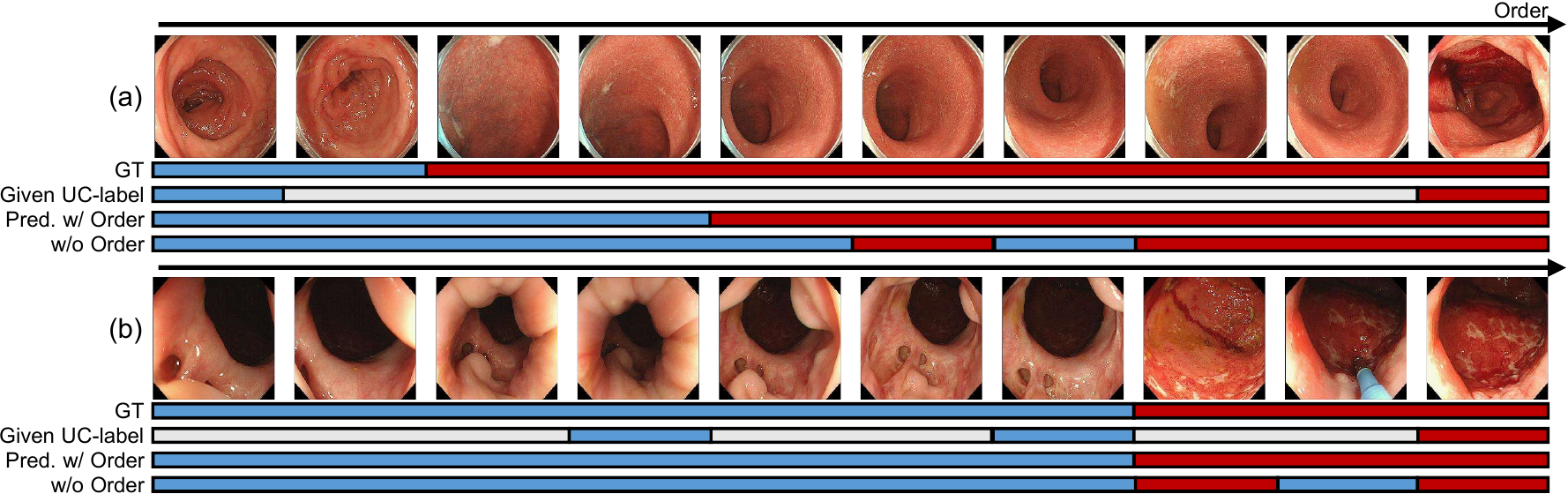}
\caption{Examples of the prediction results. Each bar represents the ground-truth labels, labels given during training, prediction results by the proposed method with and without order-guided learning. The red, blue, and gray bars represent UC, normal, and unlabeled images, respectively.}
\label{fig: example}
\end{figure}

To demonstrate the effect of the order-guided learning, the examples of prediction results were shown in Fig.~\ref{fig: example}. In this figure, the prediction results from the proposed method with the order-guided learning for temporally adjacent images tend to belong to the same class. For example, the proposed method predicted the first and second images from the right in Fig.~\ref{fig: example}(b) as the same class, whereas the proposed method without the order-guided learning predicted them as different classes.

\section{Conclusion}
We proposed a semi-supervised learning method for learning ulcerative colitis (UC) classification with limited UC-labels. The proposed method utilizes the location and temporal ordering information of endoscopic images to train the UC classifier. To obtain the features that separate the UC-dependent and location-dependent features, we introduced disentangled representation learning using location information. Moreover, to compensate for the limitation of UC-labeled data using temporal ordering information, we introduced order-guided learning, which considers the relationship between temporally adjacent images. The experimental results using endoscopic images demonstrated that the proposed method outperforms existing semi-supervised learning methods.\par

In future work, we will focus on extending the proposed framework to other tasks. Although this study applies the proposed method exclusively to UC classification, the proposed framework based on location and temporal ordering information can be applied to other tasks involving endoscopic images, such as the detection of polyps and cancer.

\subsection*{Acknowledgments}
This work was supported by JSPS KAKENHI Grant Number JP20H04211 and AMED Grant Number JP20lk1010036h0002.

\bibliography{refs}

\begin{thebibliography}{10}
\providecommand{\url}[1]{\texttt{#1}}
\providecommand{\urlprefix}{URL }
\providecommand{\doi}[1]{https://doi.org/#1}

\bibitem{Arazo_IJCNN2020}
Arazo, E., Ortego, D., Albert, P., O'Connor, N.E., McGuinness, K.:
  Pseudo-labeling and confirmation bias in deep semi-supervised learning. In:
  Proceedings of the International Joint Conference on Neural Networks (2020)

\bibitem{Wallach_NIPS2019}
Berthelot, D., Carlini, N., Goodfellow, I., Papernot, N., Oliver, A., Raffel,
  C.A.: {MixMatch}: A holistic approach to semi-supervised learning. In:
  Proceedings of the Advances in Neural Information Processing Systems. vol.~32
  (2019)

\bibitem{Cao_2020_CVPR}
Cao, K., Ji, J., Cao, Z., Chang, C.Y., Niebles, J.C.: Few-shot video
  classification via temporal alignment. In: Proceedings of the IEEE/CVF
  Conference on Computer Vision and Pattern Recognition (2020)

\bibitem{Dwibedi_CVPR2019}
Dwibedi, D., Aytar, Y., Tompson, J., Sermanet, P., Zisserman, A.: Temporal
  cycle-consistency learning. In: Proceedings of the IEEE/CVF Conference on
  Computer Vision and Pattern Recognition (2019)

\bibitem{Harada_EMBC2019}
Harada, S., Hayashi, H., Bise, R., Tanaka, K., Meng, Q., Uchida, S.: Endoscopic
  image clustering with temporal ordering information based on dynamic
  programming. In: Proceedings of the Annual International Conference of the
  IEEE Engineering in Medicine and Biology Society. pp. 3681--3684 (2019)

\bibitem{Herp2021}
Herp, J., Deding, U., Buijs, M.M., Kroijer, R., Baatrup, G., Nadimi, E.S.:
  Feature point tracking-based localization of colon capsule endoscope.
  Diagnostics  \textbf{11}(2) (2021)

\bibitem{Lee_ICML2013}
Lee, D.H.: {Pseudo-Label} : The simple and efficient semi-supervised learning
  method for deep neural networks. In: Proceedings of the ICML 2013 Workshop:
  Challenges in Representation Learning (2013)

\bibitem{Liu_NIPS2018}
Liu, A.H., Liu, Y.C., Yeh, Y.Y., Wang, Y.C.F.: A unified feature disentangler
  for multi-domain image translation and manipulation. In: Proceedings of the
  Advances in Neural Information Processing Systems. vol.~31 (2018)

\bibitem{Liu_CVPR2018}
Liu, Y., Wei, F., Shao, J., Sheng, L., Yan, J., Wang, X.: Exploring
  disentangled feature representation beyond face identification. In:
  Proceedings of the IEEE/CVF Conference on Computer Vision and Pattern
  Recognition (2018)

\bibitem{Mori2001}
Mori, K., Deguchi, D., Hasegawa, J.i., Suenaga, Y., Toriwaki, J.i., Takabatake,
  H., Natori, H.: A method for tracking the camera motion of real endoscope by
  epipolar geometry analysis and virtual endoscopy system. In: Proceedings of
  the Medical Image Computing and Computer-Assisted Intervention. pp.~1--8
  (2001)

\bibitem{Sohn_NIPS2020}
Sohn, K., Berthelot, D., Carlini, N., Zhang, Z., Zhang, H., Raffel, C.A.,
  Cubuk, E.D., Kurakin, A., Li, C.L.: {FixMatch}: Simplifying semi-supervised
  learning with consistency and confidence. In: Proceedings of the Advances in
  Neural Information Processing Systems. vol.~33, pp. 596--608 (2020)

\end{thebibliography}
\bibliographystyle{splncs04}

\end{CJK}
\end{document}